\documentclass[aps,prc,twocolumn,letterpaper,floatfix]{revtex4}
\usepackage{graphicx,amsmath,amssymb}
\usepackage[usenames,dvipsnames]{pstricks}
\usepackage{pst-grad}
\usepackage{url}

\begin{document}

\title{The Pilot-Wave Perspective on Spin}
\author{Travis Norsen}
\affiliation{Smith College \\ Northampton, MA 01060 \\ tnorsen@smith.edu}

\date{May 6, 2013}

\begin{abstract}
The alternative pilot-wave theory of quantum phenomena -- associated
especially with Louis de Broglie, David Bohm, and John Bell --
reproduces the statistical predictions of ordinary quantum mechanics,
but without recourse to special \emph{ad hoc} axioms pertaining to measurement.
That (and how) it does so is relatively straightforward to understand in
the case of position measurements and, more generally, measurements
whose outcome is ultimately registered by the position of a
pointer.   Despite a widespread belief to the contrary among
physicists, the theory can also account successfully for phenomena
involving spin.  The main goal of the paper is to explain how the pilot-wave
theory's account of spin works.  Along the way, we provide
illuminating comparisons between the orthodox and pilot-wave accounts
of spin and address some puzzles about how the pilot-wave theory relates to
the important theorems of Kochen and Specker and Bell. 
\end{abstract}

\maketitle

\section{Introduction}
\label{sec1}

In an earlier paper, which readers are encouraged to examine first and
which I refer to hereafter as ``the earlier paper,'' I
attempted to give a physicists' introduction to the alternative, de
Broglie - Bohm 
pilot-wave theory of quantum phenomena. \cite{pwp}  
As a so-called ``hidden
variable theory'' the pilot-wave theory adds something to the state
descriptions of ordinary quantum mechanics:  in addition to the usual
wave function $\Psi$ obeying the usual Schr\"odinger equation
\begin{equation}
i \hbar \frac{\partial \Psi}{\partial t} = \hat{H} \Psi
\label{sch}
\end{equation}
one also has
\emph{definite positions} for each particle in the system.  
For example, for a system of $N$ spinless non-relativistic
particles, the position $\vec{X}_n$ of the $n^{\text{th}}$ particle will
evolve according to
\begin{equation}
\frac{d \vec{X}_n(t)}{dt} = \left. \frac{\vec{j}_n(\vec{x}_1,
    \vec{x}_2, ... , \vec{x}_N, t)}{\rho(\vec{x}_1,\vec{x}_2, ...,
    \vec{x}_N, t)} \right|_{\vec{x}_k = \vec{X}_k(t) \forall k}
\label{guidance}
\end{equation}
where $\vec{j}_n = \frac{\hbar}{2mi} \left( \Psi^* \vec{\nabla}_n \Psi
  - \Psi \vec{\nabla}_n \Psi^* \right)$ is (what in ordinary QM is
termed) the probability current associated with particle $n$ and $\rho
= \Psi^* \Psi$ is (what in ordinary QM is termed) the probability
density. 

As explained in the earlier paper, the pilot-wave theory involves the
quantum equilibrium hypothesis (QEH) according to which 
the particle positions are assumed to be \emph{random}, with
initial ($t=0$) probability distribution
\begin{equation}
P[\vec{X}_1 = \vec{x}_1, ..., \vec{X}_N = \vec{x}_N] = \left|
  \Psi(\vec{x}_1, ..., \vec{x}_N, 0) \right|^2.
\end{equation}
It is then a purely mathematical consequence of Equations \eqref{sch}
and \eqref{guidance} that the probability distribution will
\emph{remain} $|\Psi|^2$-distributed for all times.  The family of
possible particle trajectories thus ``flows along with'' $\rho$, a
property that has been dubbed the ``equivariance'' of the $|\Psi|^2$
probability distribution.   \cite{qe}

Critics of the theory often argue that the addition of these definite
particle positions is pointless (or ``metaphysical'') because at the
end of the day the theory's empirical predictions match those of
ordinary quantum mechanics, which latter predictions are of course
made using the wave function alone.  What the critics fail to
appreciate, however, is that adding the particle positions allows
something to be \emph{subtracted} elsewhere in the system.  In
particular, the dynamical laws sketched above -- namely,
Equations \eqref{sch} and \eqref{guidance}  -- constitute 
the \emph{entirety} of the dynamical
postulates of the theory.  No additional
axioms or special exceptions to the usual rules -- such
as the collapse postulate of ordinary QM --  need to be introduced in
order to understand measurement or, more generally, the emergence of
the familiar everyday classical world. 

In the earlier paper, this point was developed in the context of some
simple scattering experiments involving single particles.  In such situations, the
fact that the scattered particle is found, whole, at some particular
place at the end of the experiment 
is explained in the simplest imaginable way:  there really is a
particle, following a definite trajectory and hence possessing a
perfectly definite position at all times.  The detector finds it at a
particular place (even when its wave
function is spread out across several different places) because it is
already \emph{at} a particular place before interacting with the detector. 
And equivariance, in light of the QEH, guarantees that the probability for the particle to
\emph{be at} a certain place at the end of the experiment perfectly
matches what ordinary QM would instead describe as the probability that the
measurement intervention triggers a collapse that makes
the particle suddenly materialize at that place.  
It is thus clear, in pattern, how the pilot-wave theory
reproduces the statistical predictions of ordinary QM for experiments
which end with the measurement of the position of a particle.

That the pilot-wave theory makes the same predictions as ordinary QM
for other kinds of measurements as well can be understood by including
the particles constituting the measuring device in the system under
study.  Consider for example a simple
schematic treatment of the measurement of, say, the energy of a particle
with initial wave function $\psi_0(x)$.  \cite{particle}
The measurement apparatus is imagined to have a pointer with
spatial coordinate $y$ and initial wave function $\phi(y) = \phi_0(y)$
where $\phi_0$ is a narrow packet centered at the origin, corresponding
to a ``ready'' state for the apparatus.

An interaction between the particle and the pointer is regarded
as a ``measurement of the energy
of the particle'' just in case the Schr\"odinger equation (for the
particle+pointer system) produces the following sort of time-evolution
in the case that the particle is initially in an ``energy eigenstate''
$\psi_0(x) = \psi_i(x)$ with eigenvalue $E_i$:
\begin{equation}
\psi_i(x) \phi_0(y) \rightarrow \psi_i(x) \phi_o(y-\lambda E_i).
\end{equation}
That is, at the end of the experiment, the wave function for the
pointer is now sharply peaked around some new point, $y = \lambda
E_i$, proportional to the initial energy $E_i$ of
the particle.  The pointer, in short, indicates the energy of the particle.

But the linearity of the Schr\"odinger equation then
immediately implies that, in the general case in which the particle is
initially in an arbitrary superposition of energy eigenstates, the
evolution goes as follows:
\begin{equation}
\left( \sum_i c_i \psi_i(x) \right) \phi_0(y) \rightarrow \sum_i c_i
\psi_i(x) \phi_0(y - \lambda E_i).
\end{equation}
The resulting ``Schr\"odinger cat'' type state is of course problematic from
the point of view of ordinary QM:  instead of registering some one
definite outcome for the experiment, the pointer itself ends up in an
entangled superposition.  Enter the collapse postulate to save us from
the troubling implications of universal Schr\"odinger evolution!  
But this is no problem at all in the
pilot-wave theory, according to which the (directly perceivable) final
disposition of the pointer is not to be found in the wave function,
but instead in the actual position $Y$ of the pointer particle
at the end of the experiment.
It follows from the equivariance property discussed earlier
that, assuming the initial particle positions are random in accordance
with the QEH,
there will be a probability $|c_i|^2$ that the actual
value of $Y$ at the end of the experiment is (near) $\lambda E_i$.
That is, the pointer will point to a definite place, with
probabilities given by the usual quantum rules, but, here, by
\emph{rigorously following} the basic dynamical rules, rather than
making up special exceptions when they produce embarrassing results.

One can thus understand how the pilot-wave
theory manages to reproduce the statistical predictions of ordinary
QM, at least in cases where one measures position or measures some
other quantity but by means of the position of a pointer.  But... what
about spin?  Of all the phenomena surveyed in undergraduate quantum
mechanics courses, those involving intrinsic spin and its measurement
-- which Pauli famously described as indicating a mysterious,
non-classical Zweideutigkeit or ``two-valuedness'' \cite{pauli} -- 
seem perhaps the least amenable to the pilot-wave type of analysis
just sketched.  Indeed, it is sometimes reported that the pilot-wave
theory simply cannot deal with spin phenomena in a plausible way.  \cite{motl}
This view was expressed very eloquently by one of the anonymous
referees of the earlier paper:
\begin{quote}
``It has indeed been shown that Bohmian
mechanics is equivalent to non-relativistic quantum mechanics with
respect to its predictions concerning particle positions. However the
scheme encounters major problems: in spite of a half-century of effort
by its adherents, it has not been possible to incorporate spin into
the theory in a convincing way...''   
\end{quote}
This type of view is perhaps
motivated by the various ``no hidden variables'' theorems, such as
that due to Kochen and Specker, which show quite clearly that it is mathematically
impossible to assign pre-existing values to all relevant 
spin-components of an ensemble of particles such that the quantum
mechanical predictions are reproduced.  \cite{ks,mermin}
That is, it is known to be
impossible to do, for all of the non-commuting components of spin, 
what the pilot-wave theory, as sketched above, does for position.  

And yet, in fact, it is entirely \emph{false} that the pilot-wave
theory cannot deal with spin.  Indeed, the truth is that the
pilot-wave theory deals with spin in an almost shockingly natural --
certainly a shockingly trivial -- way.  The goal of the rest of this
paper is to resolve this paradox and explain how.

The main ideas of the paper are not new.  The pilot-wave theory was
applied to spin already in 1955 by Bohm \emph{et al.} \cite{bst}, and
numerical simulations of the theory's account of spin measurements
were carried out by Dewdney \emph{et al.} in the 1980s.  \cite{dhk}
Our approach here, following Bell's more elegant treatment in Ref. \cite{bellspin},
instead aims at simplicity and accessibility.  In particular, the
proposed delta-function model
of a Stern-Gerlach experiment (the one genuine novelty in the paper) 
allows one to solve the Schr\"odinger equation
and determine the pilot-wave particle trajectories (without any
recourse to numerical simulations) using the ``plane-wave packet'' 
methods developed in the earlier paper.

That model is developed in the following section; the associated
pilot-wave particle trajectories are then displayed and analyzed in Section
\ref{sec3}.  Section \ref{sec4} explains how the pilot-wave theory
deals with cases of repeated spin measurements, while Sections
\ref{sec5} and \ref{sec6} develop examples to illustrate the
``contextuality'' and ``non-locality'' exhibited by the theory.  Some
concluding remarks about the relationship between the pilot-wave and
orthodox points of view are then made in Section \ref{sec7}.

\section{The Stern-Gerlach Experiment}
\label{sec2}

The Stern-Gerlach experiment, first and most famously performed in
1922, involves subjecting a beam of particles to an inhomogenous
magnetic field so the particles experience a force proportional to a
certain component of their intrinsic spin angular momenta. \cite{sg} 
Since, empirically, the incident beam gets split into two or more discrete
sub-beams (as opposed to a continuous distribution), the experiment is
usually understood as demonstrating the quantization of spin angular
momentum. 

\begin{figure}[t]
\begin{center}
\scalebox{.85}{
\scalebox{1} 
{
\begin{pspicture}(0,-3.14)(9.56,3.14)
\definecolor{color540}{rgb}{0.8,0.8,0.8}
\psline[linewidth=0.04cm](4.76,2.48)(4.76,0.88)
\psline[linewidth=0.04cm](4.76,0.88)(5.16,0.48)
\psline[linewidth=0.04cm](5.16,0.48)(5.56,0.88)
\psline[linewidth=0.04cm](5.56,0.88)(5.56,2.48)
\psline[linewidth=0.04cm](4.76,-1.32)(5.56,-1.32)
\psline[linewidth=0.04cm](5.56,-1.32)(5.56,-3.12)
\psline[linewidth=0.04cm](4.76,-1.32)(4.76,-3.12)
\usefont{T1}{ptm}{m}{n}
\rput(5.17,-1.815){$\text{N}$}
\psline[linewidth=0.06cm,linecolor=color540](0.96,0.08)(5.16,0.08)
\psline[linewidth=0.06cm,linecolor=color540](0.96,-0.92)(5.16,-0.92)
\psline[linewidth=0.06cm,linecolor=color540](5.16,0.08)(8.96,1.28)
\psline[linewidth=0.06cm,linecolor=color540](5.16,-0.92)(8.96,0.28)
\psline[linewidth=0.06cm,linecolor=color540](5.16,0.08)(8.96,-1.12)
\psline[linewidth=0.06cm,linecolor=color540](5.16,-0.92)(8.96,-2.12)
\usefont{T1}{ptm}{m}{n}
\rput(2.38,-0.295){$c_+ \chi_{+z} + c_- \chi_{-z}$}
\psline[linewidth=0.04cm,arrowsize=0.05291667cm 2.0,arrowlength=1.4,arrowinset=0.4]{->}(1.7,-0.62)(3.1,-0.62)
\psline[linewidth=0.04cm,arrowsize=0.05291667cm 2.0,arrowlength=1.4,arrowinset=0.4]{->}(7.24,-0.02)(8.44,0.42)
\psline[linewidth=0.04cm,arrowsize=0.05291667cm 2.0,arrowlength=1.4,arrowinset=0.4]{->}(7.22,-1.28)(8.42,-1.68)
\usefont{T1}{ptm}{m}{n}
\rput{22.869211}(0.7785742,-2.9492972){\rput(7.68,0.465){$c_+ \chi_{+z}$}}
\usefont{T1}{ptm}{m}{n}
\rput{-16.824648}(0.68174076,2.2298777){\rput(7.88,-1.175){$c_- \chi_{-z}$}}
\psline[linewidth=0.025999999cm,arrowsize=0.05291667cm 2.0,arrowlength=1.4,arrowinset=0.4]{<-}(5.16,3.12)(5.16,-0.68)
\psline[linewidth=0.025999999cm,arrowsize=0.05291667cm 2.0,arrowlength=1.4,arrowinset=0.4]{->}(4.9,-0.42)(9.3,-0.42)
\usefont{T1}{ptm}{m}{n}
\rput(9.25,-0.695){$y$}
\usefont{T1}{ptm}{m}{n}
\rput(4.82,2.945){$z$}
\psframe[linewidth=0.04,linecolor=white,dimen=outer,fillstyle=solid](5.42,1.38)(4.92,0.94)
\usefont{T1}{ptm}{m}{n}
\rput(5.14,1.185){$\text{S}$}
\end{pspicture} 
}
}
\caption{Neutral spin-1/2 particles with initial spin state $c_+
  \chi_{+z} + c_- \chi_{-z}$ move in the $y$-direction toward a
  Stern-Gerlach apparatus with magnetic field gradient along the
  z-direction.  The field delivers an impulse in the
  $\pm z$-direction to the $\chi_{\pm z}$ components of the beam, so
  downstream there arise two spatially non-overlapping sub-beams 
with amplitudes $c_+$ and $c_-$.  In ordinary quantum mechanics,
a measurement of the position of the particle downstream from the S-G
device will find the particle in (or more accurately, cause it to
materialize in) the upper/lower sub-beam with probability
$|c_\pm|^2$. 
\label{fig1}
}
\end{center}
\end{figure}
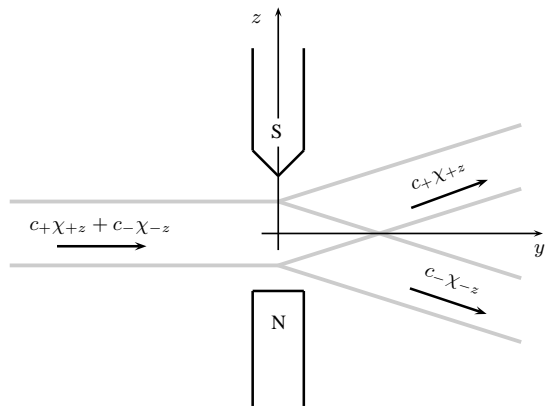

In a standard textbook treatment of the experiment, we assume a
magnetic field
\begin{equation}
\vec{B} \approx \beta z \hat{z}
\label{b}
\end{equation}
as might be produced, for example, by the magnets indicated in Figure
\ref{fig1}.  \cite{field}
For a neutral spin-1/2 particle (such as the
silver atoms used in the original S-G experiment) the interaction
Hamiltonian is
\begin{equation}
\hat{H} = - \mu \vec{\sigma} \cdot \vec{B} =  - \mu
\beta z \sigma_z
\label{h1}
\end{equation}
where $\mu$ is the magnitude of the particle's magnetic moment and
the components of $\vec{\sigma}$ are, in the usual representation, 
the Pauli matrices.  The eigenstates of this
interaction Hamiltonian will obviously be (proportional to) spinors 
$\chi_{+z} = \binom{1}{0}$ and $\chi_{-z} = \binom{0}{1}$.  
satisfying
\begin{equation}
\sigma_z \chi_{\pm z} = \pm \chi_{\pm z}.
\end{equation}
Now let us consider the spatial degrees of freedom of the particle's
wave function, focusing in particular on the dimension (here, $z$)
parallel to the magnetic field gradient.  The beam is initially
propagating, say, in the $+y$-direction, so let us assume that, in the
$z$-direction, the wave function is initially constant (over the range
where it is nonzero) and that the particles are in eigenstates of $\sigma_z$:
\begin{equation}
\Psi(z,0) = A \chi_{\pm z}.
\end{equation}
Now suppose the particles experience the magnetic field for a
finite period of time $T$ during which the interaction Hamiltonian
above dominates all other terms.  Then the Schr\"odinger equation
reads
\begin{equation}
i \hbar \frac{\partial \Psi}{\partial t} = - \mu \beta z \sigma_z \Psi
\end{equation}
which, for the assumed initial condition, has solution
\begin{equation}
\Psi(z,T)  = A e^{\pm i
  \kappa z} \chi_{\pm z} 
\end{equation}
where $\hbar \kappa = \mu \beta T$ is the magnitude of the (upward or
downward) momentum that the particle acquires as a result of
traversing the field.  For a
beam of particles initially propagating in the $y$-direction, this
impulse in the $\pm z$-direction deflects the beam either upward or
downward depending on whether the particle was initially in the spin
state $\chi_{+z}$ or $\chi_{-z}$.  Of course, in general, the initial spin
might be an arbitrary linear combination of the two z-spin
eigenstates:
\begin{equation}
\chi_0 = c_+ \chi_{+z} + c_- \chi_{-z}.
\label{spinor}
\end{equation}
But then, since Schr\"odinger's equation is linear, it is immediately
obvious that the S-G device will split the incoming wave function into
two sub-beams -- one with relative amplitude $c_+$ that is deflected
up, and one with relative amplitude $c_-$ that is deflected down.
The net effect is pictured in Figure \ref{fig1}.

For purposes of discussing the pilot-wave account of these phenomena,
it will be helpful to work with a simplified model that captures all
the physics we've just reviewed, but includes also an explicit
treatment of the particle's other spatial degrees of freedom.  This
can be done, in the spirit of Figure \ref{fig1}, by imagining that the
field $\vec{B}$ is very strong, but is non-zero only in a vanishingly
small region around $y=0$.  That is, the interaction Hamiltonian of
Equation \eqref{h1} is replaced with $\hat{H} = - \mu b z \delta(y)
\sigma_z$, so that the total Hamiltonian governing the particle's
motion in the $y-z$-plane is
\begin{equation}
\hat{H} = -\frac{\hbar^2}{2m} \frac{\partial^2}{\partial y^2} -
\frac{\hbar^2}{2m} \frac{\partial^2}{\partial z^2} - \mu b z \delta(y)
\sigma_z.
\end{equation}
Now the situation can be treated like an elementary 2D scattering
problem in wave mechanics.

To begin with, we imagine a ``plane-wave packet'' like those described
in the earlier paper:  the initial wave function should be a packet of
length $L$ along the $y$-direction and width $w$ along the
$z$-direction, with constant amplitude in the region where its
amplitude doesn't vanish, propagating initially in the
$+y$-direction with (reasonably sharply-defined) energy $E=
\hbar^2 k^2/2m$.  
Let us again start by
assuming that the incident particle's spin degrees of freedom are
described by one of the eigenspinors, $\chi_{\pm z}$.  Then, in the
time period during which the incident packet is interacting with the
field at $y=0$, the wave function can be taken to be (up to an overall
time-dependent phase which we omit for simplicity)
\begin{equation}
\Psi(y,z) = \left\{ \begin{array}{cc}
A \, e^{i k y} \, \chi_{\pm z} + B z\, e^{-iky} \, \chi_{\pm z} & \text{for } y < 0 \\
C \, e^{i k' y} e^{\pm i \kappa z} \chi_{\pm z} & \text{for } y > 0 
\end{array} \right. 
\label{wf}
\end{equation}
in the appropriate regions of the $y-z$ plane.  (Note that the factor of $z$ -- in the term,
involving $B$, representing a reflected wave -- is
put in for future convenience.)
This expression is a valid solution to the time-independent
Schr\"odinger equation for $y<0$ and $y>0$ as long as
\begin{equation}
E = \frac{\hbar^2 k^2}{2m} = \frac{\hbar^2 (k'^2 + \kappa^2)}{2m}.
\label{energyeq}
\end{equation}
In addition, the above expression should solve the Schr\"odinger
equation at $y=0$.  This will be the case if the wave function is
continuous at $y=0$
\begin{equation}
A + Bz = C \, e^{\pm i \kappa z}
\label{first}
\end{equation}
and if the following condition on the $y$-derivatives of $\psi$ (arrived at by
integrating the time-independent Schr\"odinger equation from $y=0^-$
to $y = 0^+$) is satisfied:
\begin{equation}
\left( \left. \frac{\partial \psi}{\partial y} \right|_{y=0^+} -
  \left. \frac{\partial \psi}{\partial y} \right|_{y=0^-} \right) =
\mp \frac{2m \mu b}{\hbar^2} z \psi(0,z).
\label{condition2}
\end{equation}
Plugging in our explicit expression for $\psi(y,z)$ converts Equation \eqref{condition2} into
\begin{equation}
i k' C e^{\pm i \kappa z} - i k A +i k B z = \mp \frac{2 m \mu
  b}{\hbar^2}z C \, e^{\pm i \kappa z}.
\label{second}
\end{equation}
It is obvious that these two conditions, Equations \eqref{first} and
\eqref{second}, cannot be satisfied exactly for all $z$.  However,
they can be approximately satisfied over the narrow range of $z$ where
the (width-$w$) wave packet has support.  Thus, for example, Equation
\eqref{first} is valid to first order in $\kappa w$ if $A =
C$ and
\begin{equation}
B = \pm i \kappa C.
\end{equation}
Then Equation \eqref{second} is satisfied to the same degree of
approximation if $k
\approx k'$ (which implies, in light of Equation \eqref{energyeq} that $\kappa \ll k$) and
\begin{equation}
\kappa = \frac{ m \mu b}{\hbar^2 k}.
\end{equation}
Notice that there are two distinct physical assumptions here.  First,
the impulse (of magnitude $\hbar \kappa$) imparted to the particles by
the Stern-Gerlach fields must be small compared to the initial
momentum $\hbar k$ of the particles.  Or equivalently, $\kappa / k =
\mu b / 2E$ should be small.  (Notice that $b$ has the same units as a
magnetic field even though it is not one!)  That is,  the magnetic fields should be
appropriately ``gentle''.  And second, the width $w$ of the incident
packet should be small compared to $1/\kappa$.  Then Equation
\eqref{wf} provides an acceptable description of the wave function in
the vicinity of the Stern-Gerlach apparatus while the length-$L$
packet interacts with it.  Note that, in these limits, the
($z$-dependent) amplitude of the reflected wave is (everywhere)
small.  Since it also plays no significant role in the physics to be
discussed, we therefore drop it.  (Those interested in the details
should see note \cite{reflectedwave}.)

So far we have assumed that the spin degrees of freedom are given by
one of the ($z$-direction) eigenspinors, $\chi_{\pm z}$.  But it is now
straightforward to appeal to the linearity of the
Schr\"odinger equation in order to write a solution appropriate for an
arbitrary initial spinor, like that of Equation \eqref{spinor}:
\begin{equation}
\Psi(y,z) = \left\{ \begin{array}{cc}
A \, e^{i k y} \, \left(c_+\chi_{+z} + c_- \chi_{-z} \right) & , \, y < 0 \\
A \, e^{i k' y} \left( e^{+i \kappa z} c_+ \chi_{+z} + e^{-i \kappa z}
  c_- \chi_{-z} \right) & , \,  y > 0 
\end{array} \right. 
\label{wf2}
\end{equation}
where, of course, the expression for $y<0$ applies only within the
range $-w/2 < z < w/2$ where the plane-wave packet has support, and
the expression for $y>0$ applies only within the triangular ``overlap region''
-- see Figure \ref{fig1} -- where the two (separating) components of
the wave coincide. 

\section{Particle Trajectories in the Pilot-Wave Theory}
\label{sec3}

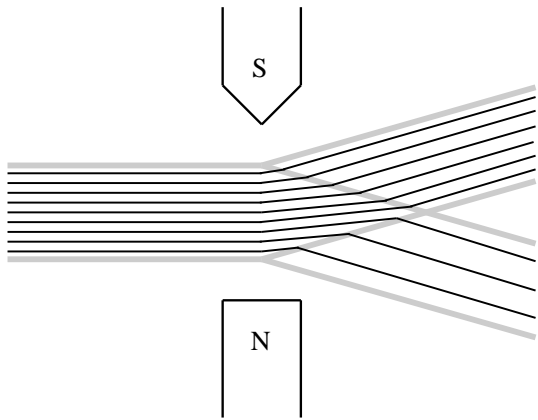
\begin{figure}[t]
\begin{center}
\scalebox{1.3}{
\scalebox{1} 
{
\begin{pspicture}(0,-2.112)(5.43,2.112)
\definecolor{color528}{rgb}{0.8,0.8,0.8}
\psline[linewidth=0.024cm](2.2,2.1)(2.2,1.3)
\psline[linewidth=0.024cm](2.2,1.3)(2.6,0.9)
\psline[linewidth=0.024cm](2.6,0.9)(3.0,1.3)
\psline[linewidth=0.024cm](3.0,1.3)(3.0,2.1)
\psline[linewidth=0.024cm](2.2,-0.9)(3.0,-0.9)
\psline[linewidth=0.024cm](3.0,-0.9)(3.0,-2.1)
\psline[linewidth=0.024cm](2.2,-0.9)(2.2,-2.1)
\usefont{T1}{ptm}{m}{n}
\rput(2.59,-1.32){\footnotesize $\text{N}$}
\psline[linewidth=0.06cm,linecolor=color528](0.0,0.48)(2.6,0.48)
\psline[linewidth=0.06cm,linecolor=color528](0.0,-0.48)(2.6,-0.48)
\psline[linewidth=0.06cm,linecolor=color528](2.6,0.48)(5.4,1.28)
\psline[linewidth=0.06cm,linecolor=color528](2.6,-0.48)(5.4,0.32)
\psline[linewidth=0.06cm,linecolor=color528](2.6,0.48)(5.4,-0.32)
\psline[linewidth=0.06cm,linecolor=color528](2.6,-0.48)(5.4,-1.28)
\psframe[linewidth=0.04,linecolor=white,dimen=outer,fillstyle=solid](2.86,1.8)(2.36,1.36)
\usefont{T1}{ptm}{m}{n}
\rput(2.58,1.48){\footnotesize $\text{S}$}
\psline[linewidth=0.02cm](2.6,0.3)(0.0,0.3)
\psline[linewidth=0.02cm](2.6,0.1)(0.0,0.1)
\psline[linewidth=0.02cm](2.6,-0.1)(0.0,-0.1)
\psline[linewidth=0.02cm](2.6,-0.3)(0.0,-0.3)
\psline[linewidth=0.02cm](2.6,0.4)(0.0,0.4)
\psline[linewidth=0.02cm](2.6,0.2)(0.0,0.2)
\psline[linewidth=0.02cm](2.6,0.0)(0.0,0.0)
\psline[linewidth=0.02cm](2.6,-0.2)(0.0,-0.2)
\psline[linewidth=0.02cm](2.6,-0.4)(0.0,-0.4)
\psline[linewidth=0.02cm](2.82,0.44)(5.4,1.18)
\psline[linewidth=0.02cm](3.32,0.28)(5.4,0.88)
\psline[linewidth=0.02cm](3.06,0.36)(5.4,1.04)
\psline[linewidth=0.02cm](3.6,0.2)(5.38,0.72)
\psline[linewidth=0.02cm](3.86,0.12)(5.4,0.58)
\psline[linewidth=0.02cm](4.12,0.06)(5.4,0.44)
\psline[linewidth=0.02cm](2.96,-0.36)(5.4,-1.08)
\psline[linewidth=0.02cm](3.48,-0.22)(5.4,-0.8)
\psline[linewidth=0.02cm](3.98,-0.06)(5.4,-0.5)
\psline[linewidth=0.02cm](2.58,0.1)(3.62,0.2)
\psline[linewidth=0.02cm](2.58,0.2)(3.34,0.28)
\psline[linewidth=0.02cm](2.58,0.3)(3.08,0.36)
\psline[linewidth=0.02cm](2.58,0.4)(2.84,0.44)
\psline[linewidth=0.02cm](2.6,0.0)(3.88,0.12)
\psline[linewidth=0.02cm](2.58,-0.1)(4.14,0.06)
\psline[linewidth=0.02cm](2.58,-0.3)(3.5,-0.22)
\psline[linewidth=0.02cm](2.6,-0.4)(2.98,-0.36)
\psline[linewidth=0.02cm](2.6,-0.2)(3.98,-0.06)
\end{pspicture} 
}
}
\caption{
Representative sample of possible particle trajectories through a
z-oriented Stern-Gerlach apparatus for an initial wave function with
spinor components $c_+ = \sqrt{2/3}$ and $c_- = \sqrt{1/3}$.  The
particle will move along with the incident wave until crossing over
into the overlap region where the $+$ and $-$ components of the wave
are beginning to separate.  In this region, the particle velocity is
the weighted average, given in Equation \eqref{overlapvel}, of the
group velocities associated with the two separating components of the
wave.  Depending on its initial lateral position within the incoming
wave packet, the particle will either be shunted into the upper ($+$)
or lower ($-$) fork.  Subsequent detection of the particle in the
upper/lower fork will result in its being identified as ``spin up/down
along $z$''.  
\label{fig2}
}
\end{center}
\end{figure}

Following the methods introduced in the earlier paper, we can use the
structure of the wave function in (especially) the ``overlap region'' to analyze
the motion of the particles in the pilot-wave theory and to
demonstrate explicitly that the theory reproduces the usual quantum
mechanical predictions for the outcome of a Stern-Gerlach spin
measurement.  The first question that must be addressed, though, is
this:  what is the analog of Equation \eqref{guidance} when the wave
function $\Psi = \binom{\Psi_+}{\Psi_-}$ is a multi-component spinor?
The answer is simply that the equation is exactly the same, except we
must sum over the spin index in both the numerator and denominator.
Or equivalently, writing $\Psi^\dagger = (\Psi^*_+ \; \Psi^*_-)$, we say
\begin{equation}
\frac{d \vec{X}(t)}{dt} = \left. \frac{\hbar}{2mi} \frac{ \Psi^\dagger
\left(  \vec{\nabla} \Psi \right) - \left( \vec{\nabla} \Psi^\dagger
\right) \Psi}{\Psi^\dagger \Psi} \right|_{\vec{x} = \vec{X}(t)} .
\label{vel}
\end{equation}
Consider, for example, the wave function $\Psi(y,z)$ in the $y<0$
region from Equation \eqref{wf2}.  We have that 
\begin{equation}
\Psi(y,z) = A e^{iky} \binom{c_+}{c_-}
\end{equation}
so that 
\begin{equation}
\Psi^\dagger (y,z) = A^* e^{-i k y} (c^*_+ \; c^*_-).
\end{equation}
Plugging into Equation \eqref{vel} then gives 
\begin{equation}
\frac{d\vec{X}}{dt} = \frac{\hbar k}{m} \hat{y}.
\end{equation}
No matter its exact location within the wave, the particle will simply 
drift in the positive $y$-direction with the group velocity of the
incident packet.  \cite{reflectedwave}

Eventually of course the particle will encounter the fields at
$y=0$ and pass over into the overlap region on the $y>0$ side. Here,
because the $+$ and $-$ components of $\Psi$ have different
dependencies on $z$, we end up with
\begin{equation}
\frac{d \vec{X}}{dt} = \frac{\hbar k'}{m} \hat{y} + \frac{\hbar
  \kappa}{m} \left( |c_+|^2 - |c_-|^2\right) \hat{z}
\label{overlapvel}
\end{equation}
which can be understood as a weighted average of the group velocities
associated with the two (now differently-directed) components of the
wave.

The family of possible particle trajectories thus looks something like
that illustrated in Figure \ref{fig2} for the particular case that
$|c_+|$ is a little bigger than $|c_-|$.  It should be clear, for
example, that if $c_+ = 1$ and $c_- = 0$, then, while in the overlap
region,  the particles will just
move with the group velocity of the $+$ component of the wave
function, and all incoming particles (no matter their lateral
position within the wave) will be shunted into the upward sub-beam.
Whereas if $c_+ = 0$ and $c_- = 1$, the particles in the overlap
region will move with the group velocity of the $-$ component and all
incoming particles will be shunted downward.  

The equivariance property mentioned in the introduction implies that
if the position of the particle within the wave is random and
$|\Psi|^2$-distributed at $t=0$, it will remain $|\Psi|^2$-distributed
for all subsequent times.  This has the immediate consequence that the
probability for the particle to be located, downstream of the
Stern-Gerlach device, in the upper (respectively, lower) sub-beam --
and hence counted as ``spin-up'' (respectively, ``spin-down'') if its
position there is detected -- is $|c_+|^2$ (respectively, $|c_-|^2$).
It is thus clear that the pilot-wave theory reproduces the usual
quantum mechanical predictions for the statistics of such spin
measurements.

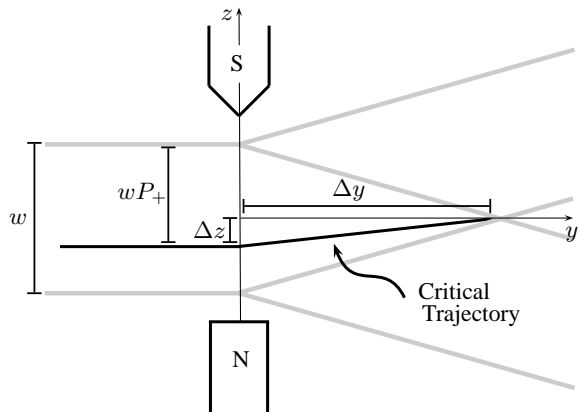
\begin{figure}[t]
\begin{center}
\scalebox{1.0}{
\scalebox{1} 
{
\begin{pspicture}(0,-2.74)(7.98,2.76)
\definecolor{color854}{rgb}{0.8,0.8,0.8}
\psline[linewidth=0.04cm](3.22,-0.52)(6.7,-0.14)
\psline[linewidth=0.06cm,linecolor=color854](3.22,0.84)(7.68,-0.42)
\psline[linewidth=0.06cm,linecolor=color854](3.24,0.84)(7.7,2.1)
\psline[linewidth=0.06cm,linecolor=color854](3.22,-1.14)(7.68,0.12)
\psline[linewidth=0.04cm](2.84,2.42)(2.84,1.62)
\psline[linewidth=0.04cm](2.84,1.62)(3.24,1.22)
\psline[linewidth=0.04cm](3.24,1.22)(3.64,1.62)
\psline[linewidth=0.04cm](3.64,1.62)(3.64,2.42)
\psline[linewidth=0.04cm](2.84,-1.52)(3.64,-1.52)
\psline[linewidth=0.04cm](3.62,-1.52)(3.62,-2.72)
\psline[linewidth=0.04cm](2.86,-1.52)(2.86,-2.72)
\usefont{T1}{ptm}{m}{n}
\rput(3.27,-2.015){$\text{N}$}
\psline[linewidth=0.06cm,linecolor=color854](0.66,0.84)(3.26,0.84)
\psline[linewidth=0.06cm,linecolor=color854](0.66,-1.14)(3.26,-1.14)
\psframe[linewidth=0.04,linecolor=white,dimen=outer,fillstyle=solid](3.5,2.12)(3.0,1.68)
\psline[linewidth=0.06cm,linecolor=color854](3.24,-1.14)(7.7,-2.4)
\usefont{T1}{ptm}{m}{n}
\rput(6.1,-1.115){Critical}
\usefont{T1}{ptm}{m}{n}
\rput(6.33,-1.435){Trajectory}
\psline[linewidth=0.04cm](0.86,-0.52)(3.26,-0.52)
\psline[linewidth=0.024cm,tbarsize=0.07055555cm 5.0]{|*-|*}(2.3,0.8)(2.3,-0.46)
\psline[linewidth=0.024cm,tbarsize=0.07055555cm 5.0]{|*-|*}(0.52,0.86)(0.52,-1.14)
\usefont{T1}{ptm}{m}{n}
\rput(0.31,-0.155){$w$}
\psline[linewidth=0.0139999995cm,arrowsize=0.05291667cm 2.0,arrowlength=1.4,arrowinset=0.4]{<-}(3.24,2.66)(3.26,-1.52)
\usefont{T1}{ptm}{m}{n}
\rput(1.96,0.185){$wP_+$}
\psline[linewidth=0.024cm,tbarsize=0.07055555cm 5.0]{|*-|*}(3.12,-0.14)(3.12,-0.46)
\usefont{T1}{ptm}{m}{n}
\rput(2.85,-0.295){$\Delta z$}
\usefont{T1}{ptm}{m}{n}
\rput(4.7,0.225){$\Delta y$}
\psline[linewidth=0.024cm,tbarsize=0.07055555cm 5.0]{|*-|*}(3.3,0.02)(6.58,0.02)
\psbezier[linewidth=0.04,arrowsize=0.05291667cm 2.0,arrowlength=1.4,arrowinset=0.4]{<-}(4.5,-0.48)(4.76,-1.18)(5.18,-0.66)(5.44,-1.2)
\usefont{T1}{ptm}{m}{n}
\rput(3.08,2.565){$z$}
\usefont{T1}{ptm}{m}{n}
\rput(7.67,-0.355){$y$}
\psframe[linewidth=0.04,linecolor=white,dimen=outer,fillstyle=solid](3.54,2.16)(2.96,1.66)
\usefont{T1}{ptm}{m}{n}
\rput(3.22,1.925){$\text{S}$}
\psline[linewidth=0.0139999995cm,arrowsize=0.05291667cm 2.0,arrowlength=1.4,arrowinset=0.4]{->}(3.26,-0.14)(7.68,-0.14)
\end{pspicture} 
}
}
\caption{
Particles which begin above the critical trajectory will be shunted
into the upper (``spin-up'') wave packet downstream of the
Stern-Gerlach magnets, while those which begin below the critical
trajectory will be shunted into the lower (``spin-down'') packet.  For
initial particle positions that are (in accordance with the QEH) 
uniformly distributed across the
lateral extent of the incident packet, the probability for a particle to emerge
with the ``spin-up'' packet will be the fraction of the lateral
width ($w$) of the packet that is above the critical trajectory -- i.e., the
distance from the critical trajectory to the top of the incident
packet is $w P_+$.  The vertical displacement $\Delta z$ of the critical
trajectory  as it traverses the triangular overlap region is then
$wP_+ - w/2$.  By considering the right triangle that is the lower
half of the overlap region (and whose hypotenuse is one edge of the
``spin-up'' packet), one can see that the horizontal extent of the
overlap region, $\Delta y$, is given by $\frac{w}{2} \frac{\kappa}{k'}$:
the slope, $(w/2)/\Delta y$, of the hypotenuse should match the
ratio $\kappa / k'$ of the $z$- and $y$-components of the group
velocity associated with the  ``spin-up'' part of the wave. 
\label{fig3}
}
\end{center}
\end{figure}

One of the nice things about the way we have set things up, though, is
that this claim can be demonstrated explicitly by considering the
``critical trajectory'' \cite{bh} which divides the trajectories emerging into
the upper and lower beams.  By definition, the criticial trajectory
arrives just at the vertex on the right of the triangular overlap
region behind the magnets, as shown in Figure \ref{fig3}. 
The slope of the criticial trajectory is given by the ratio of the
$z-$ and $y-$components of the velocity in the overlap region,
Equation \eqref{overlapvel}:
\begin{equation}
\text{slope} = \frac{|c_+|^2 - |c_-|^2}{|c_+|^2+|c_-|^2}
\frac{\kappa}{k'}.
\end{equation}
On the other hand, as explained in the Figure caption, the critical
trajectory traverses a distance $\Delta y = \frac{w k'}{2 \kappa}$ in
the $y-$direction and a distance $\Delta z = wP_+ - w/2$ in the
$z-$direction, where $P_+$ is the fraction of the lateral width $w$ of
the beam with the property that, should the particle begin there, it
wil end up going into the upper sub-beam.  In short, $P_+$
represents the probability that the S-G spin measurement will yield
the outcome ``spin-up''.   Setting
\begin{equation}
\text{slope} = \frac{\Delta z}{\Delta y}
\end{equation}
and solving for $P_+$ one indeed recovers that
\begin{equation}
P_+ = |c_+|^2
\end{equation}
confirming explicitly that, indeed, the pilot-wave dynamics for the
wave and particle are compatible in the way expressed by the formal
equivariance property:  the particle trajectories ``bend just the
right amount'' to yield a final probability distribution for the
particle position that is consistent with the usual QM predictions
and, more importantly, what is seen in experiments.

\section{Additional Measurements}
\label{sec4}

So far we have concentrated on the measurement of the $z$-component of
the spin using a Stern-Gerlach device whose magnetic field gradient is
along the $z$-direction.  By simply rotating the device, however, we can
also measure the component of the spin along (say) an arbitrary
direction $\hat{n}$ in the $x-z$-plane:
\begin{equation}
\hat{n} = \cos\theta \hat{z} + \sin \theta \hat{x}.
\end{equation}
The whole measurement procedure can be analyzed in exactly the same
way we've done above, \emph{mutatis mutandis}:  the eigenspinors
$\chi_{\pm n}$ will now be the eigenvectors of 
\begin{equation}
\sigma_n = \left( \begin{array}{cc} 
\cos \theta & \sin \theta \\
\sin \theta & -\cos \theta \end{array} \right),
\end{equation}
an arbitrary initial wave-packet (proportional to initial spinor
$\chi_0$) can be written as a linear combination of these
eigen-spinors
\begin{equation}
\chi_0 = c_{+n} \chi_{+n} + c_{-n} \chi_{-n}
\end{equation}
and the whole analysis of Sections II and III goes through, with $z$
being simply everywhere replaced by $n$.  \cite{angle}
So it should be immediately
clear that the pilot wave theory reproduces the usual quantum
statistical predictions for spin measurements, whether it is the $z$-
or some other component of spin that is to be measured.  

Perhaps the theory will fail, however, when we consider the
possibility of \emph{subsequent} and/or \emph{repeated} measurements?
The pilot-wave theory is, after all, a deterministic hidden variable
theory:  for a given experimental setup, 
the outcome of a spin measurement is determined by
the initial state of the ``particle'' (meaning here the particle+wave 
combination).  Standard textbook discussions might perhaps suggest
that there would be a problem.

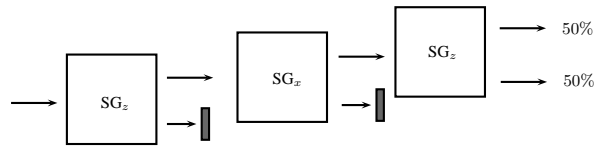
\begin{figure}[t]
\begin{center}
\scalebox{0.7}{
\scalebox{1} 
{
\begin{pspicture}(0,-1.34)(11.34,1.34)
\definecolor{color79b}{rgb}{0.4,0.4,0.4}
\psframe[linewidth=0.04,dimen=outer](2.78,0.44)(1.04,-1.3)
\psline[linewidth=0.04cm,arrowsize=0.05291667cm 2.0,arrowlength=1.4,arrowinset=0.4]{->}(0.0,-0.5)(0.88,-0.5)
\usefont{T1}{ptm}{m}{n}
\rput(1.98,-0.515){$\text{SG}_z$}
\psline[linewidth=0.04cm,arrowsize=0.05291667cm 2.0,arrowlength=1.4,arrowinset=0.4]{->}(2.96,-0.9)(3.52,-0.9)
\psframe[linewidth=0.04,dimen=outer,fillstyle=solid,fillcolor=color79b](3.78,-0.58)(3.6,-1.22)
\psline[linewidth=0.04cm,arrowsize=0.05291667cm 2.0,arrowlength=1.4,arrowinset=0.4]{->}(2.96,-0.02)(3.84,-0.02)
\psframe[linewidth=0.04,dimen=outer](6.02,0.86)(4.28,-0.88)
\usefont{T1}{ptm}{m}{n}
\rput(5.25,-0.055){$\text{SG}_x$}
\psline[linewidth=0.04cm,arrowsize=0.05291667cm 2.0,arrowlength=1.4,arrowinset=0.4]{->}(6.22,0.38)(7.1,0.38)
\psline[linewidth=0.04cm,arrowsize=0.05291667cm 2.0,arrowlength=1.4,arrowinset=0.4]{->}(6.28,-0.54)(6.84,-0.54)
\psframe[linewidth=0.04,dimen=outer,fillstyle=solid,fillcolor=color79b](7.1,-0.22)(6.92,-0.86)
\psframe[linewidth=0.04,dimen=outer](9.02,1.34)(7.28,-0.4)
\usefont{T1}{ptm}{m}{n}
\rput(8.2,0.465){$\text{SG}_z$}
\psline[linewidth=0.04cm,arrowsize=0.05291667cm 2.0,arrowlength=1.4,arrowinset=0.4]{->}(9.28,0.92)(10.16,0.92)
\psline[linewidth=0.04cm,arrowsize=0.05291667cm 2.0,arrowlength=1.4,arrowinset=0.4]{->}(9.3,-0.1)(10.18,-0.1)
\usefont{T1}{ptm}{m}{n}
\rput(10.77,0.925){$50\%$}
\usefont{T1}{ptm}{m}{n}
\rput(10.79,-0.075){$50\%$}
\end{pspicture} 
}
}
\caption{
Schematic representation of a series of Stern-Gerlach spin
measurements, in which the devices are replaced by ``black boxes''
with two output ports, one for ``spin-up'' along the direction $n$
indicated by the box's label ``SG$_{n}$'', and one for ``spin-down''.
Here, particles which emerge as ``spin-up'' along the $z$-direction
and then ``spin-up'' along the $x$-direction are subjected to a
further measurement of $z$-spin.  Half of the particles exiting this
final SG$_z$ device are ``spin-up'', with the other half being
``spin-down''.  As discussed in the text, both ordinary QM and 
the pilot-wave theory are able to
account for the observed statistics, because in both theories the
state of a particle is in general \emph{affected} by subjecting it to
a spin measurement.  In particular, the state of a particle
exiting the first SG$_z$ device is not the same as the state of that
same particle when it exits the intermediate SG$_x$ device.  
\label{fig4}
}
\end{center}
\end{figure}

For example, suppose that particles emerging from the ``spin-up'' port
of a $z$-oriented Stern-Gerlach device (SG$_z$) are subsequently sent through
an  SG$_x$ device; those particles emerging as
``spin-up'' along $x$ are then subjected to a further measurement of
their spin's $z$-component.  (See Figure \ref{fig4}.)  
The result, of course, is that fully half
of the particles entering it will emerge from the final SG$_z$ device
as ``spin-down''.  The usual interpretation is that spin-along-$z$ and
spin-along-$x$ are incompatible properties, corresponding to
non-commuting quantum mechanical operators, so the determination of a
definite value for spin-along-$x$ completely erases the
previously-definite value for spin-along-$z$.  Surely a non-quantum,
deterministic hidden variable theory could not reproduce this
paradigmatically quantum result?

But a little thought reveals that, yes, the pilot-wave theory
reproduces this prediction quite trivially.  The crucial point is
that, although the particles \emph{entering} the SG$_x$ device were
being guided by waves proportional to $\binom{1}{0}$, 
the wave guiding those particles which successfully \emph{emerge} from the
``spin-up'' port of the SG$_x$ device is proportional to $\chi_{+x} =
\frac{1}{\sqrt{2}} \binom{1}{1}$.  The $\chi_{+x}$ and $\chi_{-x}$ 
components of the wave get
separated by the SG$_x$ device, whereas the particle goes one way or the
other.  And, by definition, if the particle ends up in the
``spin-up-along-$x$'' beam, downstream of the SG$_x$ device, the
component of the wave now surrounding and guiding it, is the
``spin-up-along-$x$'' part.  So it is as if the particle's quantum state
has ``collapsed'', as per the usual quantum mechanical rules,
even though, in fact, no collapse -- no violation of the usual
Schr\"odinger evolution of the wave -- has occurred.  The other,
``spin-down-along-$x$'', part of the wave is still present -- it's
just ``over there'', far away from the actual location of the
particle, and hence dynamically irrelevant to the present or future
motion of the particle. \cite{future}  This separation of the different spin
components of the wave packet of course happens in ordinary quantum
mechanics as well.  But in ordinary QM, there is nothing like the
\emph{actual} position of the particle in addition to the wave
function, to warrant talk of the particle \emph{actually} being up
there (i.e., talk of the spin measurement having actually had the
outcome ``spin-up''), and so additional dynamical postulates are
required.  This is an illustration of the point made in the
introduction, that although the pilot-wave theory adds something to
the physical state descriptions, it is not really more (let alone
pointlessly more) complicated than ordinary QM, since this addition
allows for the subtraction of the otherwise rather dubious measurement
axioms.



What is actually showed by this kind of example is not that one
cannot explain the observed statistics of spin measurements with a
non-quantum, deterministic, hidden variable theory, but rather only
that the physical state of a particle (known, say, to be
``spin-up-along-$z$'') must be \emph{affected} by a measurement of
(for example) its spin-along-$x$.  Ordinary quantum theory of
course exhibits this behavior:  in quantum
mechanics, the state of the particle, after a measurement, is
postulated to ``collapse'' to the eigenstate of the appropriate
operator corresponding to the actually-observed outcome.  (It is worth
noting here that the ``actually-observed outcome'' is yet
\emph{another} additional postulate of the theory -- for Bohr, for
example, this was realized in some classical macroscopic pointer
somewhere, whose \emph{physical} relationship to the particle in
question is, at best, obscure.)  The point here is that the pilot-wave
theory is also exhibits this behavior:  the physical state
of (in particular) the wave guiding the particle is different  before,
and after, the intermediate SG$_x$ device.  But in the pilot-wave
theory, this difference -- the physical influence of the measurement
on the properties of the system in question -- is a natural and
straightforward consequence of the \emph{usual} -- the
\emph{universal}, the \emph{exceptionless} -- dynamical laws.

\section{Contextuality}
\label{sec5}

In the previous section, I stressed that the pilot-wave
theory is not the ``naive'' sort of hidden-variable theory that is
sometimes discussed (and, when discussed, always refuted with great
pomp!) in textbooks.  In these ``naive'' theories, measurements simply
passively reveal  some pre-existing value of the property being
measured, without affecting the state of the particle.  In the
pilot-wave theory, by contrast, there is a natural mechanism (which,
quite literally, is nothing but the Equations, \eqref{sch} and
\eqref{guidance}, defining the theory's dynamics) whereby
the physical measurement intervention affects the state of the
measured system.  There is thus a sense in which, for the pilot-wave
theory just as for ordinary QM, the measurement cannot be thought of
as passively revealing some pre-existing quantity, but should instead
be thought of as an active intervention which brings about the new, final
state of the particle corresponding to the measurement outcome.

But there is an even deeper sense in which, for the
pilot-wave theory, the measurement cannot be thought of as passively 
revealing a
pre-existing value.  Let us discuss this in terms of a concrete
example.  Recall the SG$_z$ device sketched in Figure \ref{fig1}:  the
magnets produce a magnetic field near the origin which can be
approximated by the expression in Equation \eqref{b}.  The field is
such that a \emph{classical} particle, with magnetic dipole moment in
the $+z$ direction, will feel a force in the $+z$ direction and hence
be deflected up upon passing through the SG device.  Likewise, a
classical particle with magnetic dipole moment in the $-z$ direction
will feel a force in the $-z$ direction and will hence be deflected
down.  This is, in essence, the basis for saying that, when a particle
emerges from the SG device having been deflected up, it is ``spin-up
along $z$'', etc.  

\begin{figure}[t]
\begin{center}
\scalebox{0.85}{
\scalebox{1} 
{
\begin{pspicture}(0,-3.14)(9.56,3.14)
\definecolor{color39}{rgb}{0.8,0.8,0.8}
\psline[linewidth=0.04cm](4.76,2.48)(4.76,0.88)
\psline[linewidth=0.04cm](4.76,0.88)(5.16,0.48)
\psline[linewidth=0.04cm](5.16,0.48)(5.56,0.88)
\psline[linewidth=0.04cm](5.56,0.88)(5.56,2.48)
\psline[linewidth=0.04cm](4.76,-1.32)(5.56,-1.32)
\psline[linewidth=0.04cm](5.56,-1.32)(5.56,-3.12)
\psline[linewidth=0.04cm](4.76,-1.32)(4.76,-3.12)
\usefont{T1}{ptm}{m}{n}
\rput(5.14,-1.815){$\text{S}$}
\psline[linewidth=0.06cm,linecolor=color39](0.96,0.08)(5.16,0.08)
\psline[linewidth=0.06cm,linecolor=color39](0.96,-0.92)(5.16,-0.92)
\psline[linewidth=0.06cm,linecolor=color39](5.16,0.08)(8.96,1.28)
\psline[linewidth=0.06cm,linecolor=color39](5.16,-0.92)(8.96,0.28)
\psline[linewidth=0.06cm,linecolor=color39](5.16,0.08)(8.96,-1.12)
\psline[linewidth=0.06cm,linecolor=color39](5.16,-0.92)(8.96,-2.12)
\usefont{T1}{ptm}{m}{n}
\rput(2.33,-0.295){$c_+ \chi_{+z}+ c_- \chi_{-z}$}
\psline[linewidth=0.04cm,arrowsize=0.05291667cm 2.0,arrowlength=1.4,arrowinset=0.4]{->}(1.7,-0.62)(3.1,-0.62)
\psline[linewidth=0.04cm,arrowsize=0.05291667cm 2.0,arrowlength=1.4,arrowinset=0.4]{->}(7.24,-0.02)(8.44,0.42)
\psline[linewidth=0.04cm,arrowsize=0.05291667cm 2.0,arrowlength=1.4,arrowinset=0.4]{->}(7.22,-1.28)(8.42,-1.68)
\usefont{T1}{ptm}{m}{n}
\rput{20.382967}(0.64262,-2.6745803){\rput(7.76,0.465){$c_- \chi_{-z}$}}
\usefont{T1}{ptm}{m}{n}
\rput{-16.824648}(0.6727313,2.2489564){\rput(7.94,-1.135){$c_+ \chi_{+z}$}}
\psline[linewidth=0.025999999cm,arrowsize=0.05291667cm 2.0,arrowlength=1.4,arrowinset=0.4]{<-}(5.16,3.12)(5.16,-0.68)
\psline[linewidth=0.025999999cm,arrowsize=0.05291667cm 2.0,arrowlength=1.4,arrowinset=0.4]{->}(4.9,-0.42)(9.3,-0.42)
\usefont{T1}{ptm}{m}{n}
\rput(9.25,-0.695){$y$}
\usefont{T1}{ptm}{m}{n}
\rput(4.82,2.945){$z$}
\psframe[linewidth=0.04,linecolor=white,dimen=outer,fillstyle=solid](5.42,1.38)(4.92,0.94)
\usefont{T1}{ptm}{m}{n}
\rput(5.17,1.185){$\text{N}$}
\end{pspicture} 
}
}
\caption{
An alternative way to ``measure the $z$-component of the spin of a
particle'' is to pass the particle through the $\text{SG}'_z$ device
shown here.
This is the same as an $\text{SG}_z$ device, but with the polarity of
the magnets -- and hence the calibration of the device -- reversed:
particles that deflect \emph{up} are now counted as ``spin-down along
$z$'', while particles that deflect \emph{down} are now counted as
``spin-up along $z$''.  
\label{fig5}
}
\end{center}
\end{figure}
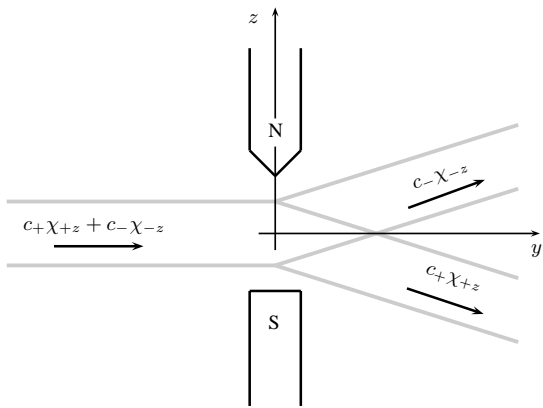

But other ways of measuring the $z$-component of a particle's spin can
also be contemplated.  For example, imagine a device -- let us call it
here an $\text{SG}'_z$ device -- like that indicated in Figure
\ref{fig5}.  The structure is identical to the original $\text{SG}_z$
device, except that the polarity of the magnets has been reversed and
hence Equation \eqref{b} is replaced with
\begin{equation}
\vec{B}' \approx - \beta z \hat{z}.
\end{equation}
This reversal of the magnetic field gradient reverses the effect on
particles passing through the device:
now, a classical particle with a magnetic dipole
moment in the $+z$ direction will feel a force in the $-z$ direction
and hence be deflected \emph{down} upon passing through the device,
while a classical particle with magnetic dipole moment in the $-z$
direction will feel a force in the $+z$ direction and will hence be
deflected \emph{up}.  The $\text{SG}'_z$ device is a perfectly valid
one for measuring the $z$-component of the spin of a particle; it
merely has a different ``calibration'' (one might say) than the
original device.  Whereas, with the original device, particles
deflected \emph{up} are declared to be ``spin-up along $z$'',
particles deflected up by the modifed device are instead declared to
be ``spin-\emph{down} along $z$'', and vice versa.  And indeed, a full
quantum mechanical analysis, parallel to that undertaken for the
original $\text{SG}_z$ device in Section \ref{sec3}, leads (in a
rather obvious and trivial way) to the conclusion that an incident
wave packet proportional to $c_+ \chi_{+z} + c_- \chi_{-z}$ will be split,
upon passage through the $\text{SG}'_z$ device, into two sub-beams,
one proportional to $c_+ \chi_{+z}$ that has been deflected \emph{down}
and the other proportional to $c_- \chi_{-z}$ that has been deflected
\emph{up}.

The significance of these two distinct pieces of experimental
apparatus -- both equally qualified to ``measure the $z$-component of
the spin of a particle'' -- becomes clear when we consider what
happens when, for example, a particle with initial wave function
proportional to
$\frac{1}{\sqrt{2}} \binom{1}{1}$ is incident on the two devices.  For
this case, the particle velocity in the overlap region is proportional
to $\hat{y}$, i.e., the particle trajectories simply continue in a
straight line all the way through the overlap region until they
finally emerge into one of the now spatially-separated sub-beams
corresponding to their initial lateral positions within the incident
packet.  That part is identical, whether the original $\text{SG}_z$,
or instead the alternative $\text{SG}'_z$, device is involved.  
But there is a crucial difference between the two devices:  if (say)
the particle happens to start in the upper half of the incident
packet, it is destined to eventually find itself in the
upward-deflected sub-beam and hence be counted as ``spin-up along
$z$'' -- if the measurement is carried out using the original
$\text{SG}_z$ device.  But if instead the measurement is
carried out using the alternative $\text{SG}'_z$ device, the exact
same ``particle'' -- that is, the exact same wave function and the exact
same particle location in the upper half of the wave packet -- is
instead destined to eventually find itself in the upward-deflected
sub-beam and hence be counted as ``spin-\emph{down} along $z$''.  
\cite{context}


In short, for the (deterministic, hidden variable) pilot-wave theory, 
the outcome of ``measuring the $z$-component of the spin of the
particle'' is \emph{not} simply a function of the initial state of the
``particle'' (i.e., the particle+wave complex).  The exact same
initial state can yield either of the two possible
measurement outcomes, depending on which of two possible experimental
devices is chosen for performing the experiment.  

This is a concrete illustration of the fact that is usually put as
follows:  for the pilot-wave theory, spin is ``contextual''.  That is,
the outcome of a measurement of a certain component of a particle's
spin depends, in the pilot-wave theory, not just on the initial state
of the particle but also on the overall experimental context -- i.e.,
on the particular ``way'' that the measurement is implemented.  

This ``contextuality'' should be contrasted to the ``non-contextual''
type of hidden variable theory in which, for each (Hermitian) 
quantum mechanical operator (corresponding to some ``observable''
property), each particle in an ensemble possesses a definite value
which is simply revealed by the appropriate kind of experiment,
independent of which specific experimental implementation is used to
measure the observable in question.  In particular,  a
non-contextual hidden variable theory would (by definition) assign a
definite value to the $z$-component of each particle's spin,
independent of whether the spin is measured using an $\text{SG}_z$ or
an $\text{SG}'_z$ device.  More generally, non-contextuality can be
understood as the requirement that each observable should possess a
particular definite value, that will be revealed by a measurement of
that observable, \emph{no matter what compatible observables are
  perhaps being measured simultaneously}.  (See the following section
for a concrete example of this more general sort of contextuality.)

That the ``no hidden variables'' theorems of von Neumann, Jauch-Piron,
and Gleason tacitly assume non-contextuality (or an even stronger and
less innocent requirement) was first clearly pointed out in Bell's
landmark 1966 paper, ``On the Problem of Hidden Variables in Quantum
Mechanics.''  \cite{bell66}  (The paper, incidentally, was written in
1964, prior to Bell's more famous 1964 paper proving what is now
called ``Bell's Theorem'', but remained unpublished until 1966 
due to an editorial
accident.)  The somewhat more famous Kochen-Specker ``no hidden
variables'' theorem, which appeared the year after Bell's paper, also
assumes non-contextuality, as discussed in the beautiful review
paper by Mermin.  \cite{ks, mermin}

As Bell explains, these proofs all rely on relating ``in a nontrivial
way the results of experiments which cannot be performed
simultaneously.''  (For example, they assume that the results of an 
$\text{SG}_z$-based and an $\text{SG}'_z$-based measurement should be
the same.)  Bell elaborates:
\begin{quote}
``It was tacitly assumed that
measurement of an observable must yield the same value independently
of what other measurements may be made simultaneously.  Thus as well
as [some observable with corresponding QM operator $\hat{A}$] say, one
might measure \emph{either} [$\hat{B}$] \emph{or} [$\hat{C}$], where
[$\hat{B}$ and $\hat{C}$ commute with $\hat{A}$, but not necessarily with each
other].  These different 
possibilities require different experimental arrangements;
there is no \emph{a priori} reason to believe that  the results for
[$\hat{A}$] should be the same.  The result of an observation may
reasonably depend not only on the state of the system (including
hidden variables) but also on the complete disposition of the
apparatus...'' \cite{bell66}
\end{quote}
Bell then references Bohr's insistence on remembering ``the
impossibility of any sharp distinction between the behavior of atomic
objects and the interaction with the measuring instruments which serve
to define the conditions under which the phenomena appear.''
\cite{bohr}  Abner Shimony has aptly described Bell's invocation of
Bohr (the arch-opponent of hidden variables) in Bell's \emph{defense}
of hidden variables (against the theorems supposedly, but not
actually, showing them to be impossible) as a ``judo-like
manoeuvre''.  \cite{shimony}

For our purposes, the upshot of all this is the following.  In
the literature on hidden variables, one often finds the
implication that the requirement of non-contextuality is quite
reasonable, in the sense that contextuality would supposedly involve
putting in ``by hand'' an obviously
implausible \emph{ad hoc} kind of apparatus-dependence.  But as the simple
example of the contextuality of the pilot-wave theory (involving the
distinct outcomes for $\text{SG}_z$- and $\text{SG}'_z$-based
measurements of $z$-spin) makes clear, this is not the case at all.
The ``context-dependence'' arises in a perfectly straightforward and
natural way, again as a direct consequence of the fundamental
dynamical postulates of the theory, Equations \eqref{sch} and
\eqref{guidance}.

\section{Non-Locality}
\label{sec6}



Bell's 1966 paper closes by noting that Bohm's 1952 pilot-wave
theory eludes the various ``no hidden variables'' theorems by means of
its contextuality -- but that the context-dependence implies a
non-local action at a distance in situations involving entangled but
spatially-separated 
particles:  ``in this theory an explicit causal mechanism exists
whereby the disposition of one piece of apparatus affects the results
obtained with a distant piece.''  \cite{bell66}  Let us explore this type of
situation, in which the pilot-wave theory's contextuality implies a
non-local dependence on a remote context. 

Consider then a pair of spin-$1/2$ particles, propagating in opposite
directions along (and already widely separated along) the $y$-axis.
The initial spatial part of the 2-particle wave function can be taken
to be
\begin{equation}
\psi(\vec{x}_1, \vec{x}_2) = \Phi^{(0,d,0)}_k (\vec{x}_1) \Phi^{(0,-d,0)}_{-k} (\vec{x}_2)
\end{equation}
where $\Phi^{(0,d,0)}_k(\vec{x})$ represents a plane-wave packet \cite{pwp}
centered at $\vec{x} = (0,d,0)$ with central wave vector
$\vec{k} = k \hat{y}$.  Consider first the case in which
the spin part of the initial state also factorizes, so that the total
state can be written
\begin{equation}
\Psi(\vec{x}_1,\vec{x}_2) = \left( \Phi^{(0,d,0)}_k (\vec{x}_1) \chi^1_0\right)
\left( \Phi^{(0,-d,0)}_k(\vec{x}_2) \chi^2_0 \right).
\label{factor}
\end{equation}
We may introduce the so-called ``conditional wave function'' (CWF)
for particle 1 as follows:
\begin{equation}
\Psi_1(\vec{x}) = \left. \Psi(\vec{x}, \vec{x}_2) \right|_{ \vec{x}_2
  = \vec{X}_2}
\end{equation}
where $\vec{X}_2$ is, of course, the actual position of particle 2.
The CWF for particle 2 is, obviously, defined in a parallel way.

These 
CWFs are convenient because the dynamical law for the particle
positions, Equation \eqref{guidance}, can be reformulated to give, for
each particle, an expression for the particle velocity in terms of its
own CWF:
\begin{equation}
\frac{d \vec{X}_n}{dt} = \left. \frac{\hbar}{2mi} \frac{
    \Psi_n^\dagger (\vec{\nabla} \Psi_n ) - (\vec{\nabla}
    \Psi_n^\dagger) \Psi_n }{\Psi_n^\dagger \Psi_n} \right|_{\vec{x} = \vec{X}_n}.
\label{CWFg}
\end{equation}
The CWFs can thus be thought of as ``one-particle wave functions'',
each of which guides its corresponding particle, in exactly the way we
have discussed previously for the pilot-wave theory of single particles.

The CWFs have however several perhaps-unexpected properties that should be
acknowledged.  First, it should be appreciated that,
in general, the CWFs do not obey some simple one-particle
Schr\"odinger equation.  \cite{telb}   Evaluating Equation
\eqref{factor} at $\vec{x}_2 = \vec{X}_2$ gives an overall
(time-dependent) multiplicative factor which means that (even in the
case of non-entangled particles) the overall phase and normalization
of the CWF can change erratically.  This, however, is irrelevant to
the motion of the associated particle, since any multiplicative factor
simply cancels out in Equation \eqref{CWFg}.  In addition, the CWF
of \emph{each} particle carries spin indices for \emph{both}
particles.  In the case where the
two-particle wave function is factorizable, as in Equation
\eqref{factor}, the particle 2 spinor ($\chi^2_0$) is irrelevant to
the motion of particle 1, and vice versa.  The ``extra'' spinor thus
acts just like the overall multiplicative constant.  It is thus clear
that when the two-particle wave function factorizes -- when the two
particles are not entangled --  the motion of the two particles will be fully independent:  if
the particles subsequently encounter Stern-Gerlach devices, each will
(independently and separately) act exactly as described in the
previous sections.  

If, however, the spins of the two particles are initially entangled, as for
example in
\begin{equation}
\Psi(\vec{x}_1,\vec{x}_2) = \psi(\vec{x}_1,\vec{x}_2)
\frac{1}{\sqrt{2}} \left( \chi^1_{+z}\chi^2_{-z} - \chi^1_{-z} \chi^2_{+z}\right)
\label{ent}
\end{equation}
then the non-locality appears.  Suppose for example that both
particles are to be subjected to $\text{SG}_z$-based measurements of their
$z$-spins.  And suppose that particle 1 encounters its
$\text{SG}_z$ device first.  A simple calculation shows that, in the
overlap region behind the magnets, the velocity of particle 1 will be
proportional to $\hat{y}$ (i.e., it will simply continue in a straight
horizontal line through the overlap region).  It will then exit the
overlap region into one or the other of the two downstream sub-beams,
depending on its initial $z$-coordinate:  if
the particle happens to have started in the upper half of the packet
it will end up deflecting up and being counted as ``spin-up'', whereas
if it happens to have started in the lower half of the packet it will
end up deflecting down and being counted as ``spin-down''.
Supposing, for notational simplicity, that the two now
spatially-separated sub-beams are bent (say, by additional appropriate
SG type magnets) so that they again propagate in the $+y$-direction
but with displacements $\pm \Delta$ in the $z$-direction \cite{delta}, the
two-particle wave function after particle 1 has passed through its
$\text{SG}_z$ device will take the form:
\begin{eqnarray}
\Psi \! &=& \! \frac{1}{\sqrt{2}} \left[ 
\Phi_k^{(0,d',\Delta)} (\vec{x}_1) \, \Phi_{-k}^{(0,-d',0)}
(\vec{x}_2) \, \chi^1_{+z} \chi^2_{-z} \right. \\
&& \; \; \; \; \; \; \; \left. - \; \Phi_k^{(0,d',-\Delta)} (\vec{x}_1) \,
  \Phi_{-k}^{(0,-d',0)} (\vec{x}_2) \, \chi^1_{-z} \chi^2_{+z} \right]. \nonumber
\end{eqnarray}
That is, the overall wave function is a superposition of two terms --
one, proportional to $\chi^1_{+z} \chi^2_{-z}$, in which the particle 1 wave
packet has been displaced a distance $\Delta$ in the \emph{positive}
$z$-direction, and the other, proportional to $\chi^1_{-z} \chi^2_{+z}$, in
which the particle 1 wave packet has been displaced a distance
$\Delta$ in the \emph{negative} $z$-direction.

The actual location $\vec{X}_1$ of particle 1 will be \emph{either}
near $z=\Delta$ or near $z = -\Delta$ (again, depending on its random
initial position).  It is thus meaningful already at this stage to
speak of the actual outcome of the $\text{SG}_z$-based measurement of
the $z$-spin of particle 1.  The crucial point is now that the CWF for
particle 2 -- the thing that will determine how particle 2 behaves
when it subsequently encounters its $\text{SG}_z$ device -- depends on
where particle 1 ended up.  If particle 1 went up (that is,
if $\vec{X}_1$ is now in the support of $\Phi_k^{(0,d',\Delta)}$) then
the CWF of particle 2 is (proportional to)
\begin{equation}
\Psi_2(\vec{x}) = \Phi_{-k}^{(0,-d',0)}(\vec{x}) \, \chi_{-z}
\end{equation}
and its subsequent interaction with an $\text{SG}_z$ device will,
independent of the exact position $\vec{X}_2$, result in its being
deflected \emph{down}.  
Whereas if instead particle 1 ended up going down (that is, if $\vec{X}_2$
is now in the support of $\Phi_k^{(0,d',-\Delta)}$) then the CWF of
particle 2 is instead (proportional to)
\begin{equation}
\Psi_2(\vec{x}) = \Phi_{-k}^{(0,-d',0)} (\vec{x}) \, \chi_{+z}
\end{equation}
and its subsequent interaction with an $\text{SG}_z$ device will,
independent of the exact position $\vec{X}_2$, instead result in its
being deflected \emph{up}.  That is, the CWF for particle 2
\emph{collapses}, as a result of the measurement on particle 1,
to a state of definite $z$-spin.  This
happens, however, despite the fact that the two-particle wave function
obeys the (unitary) Schr\"odinger equation without exception!  And the
usual quantum mechanical statistics are reproduced:  the
two possible joint outcomes (particle 1 is spin-up and particle 2 is
spin-down, or particle 1 is spin-down and particle 2 is spin-up) each
occur with 50\% probability.   Of course, the pilot-wave theory being
after all deterministic, these probabilities are quite reducible.  In
particular, which of the two joint outcomes occurs depends on the
random initial position of particle 1 (specifically, its initial
$z$-coordinate).

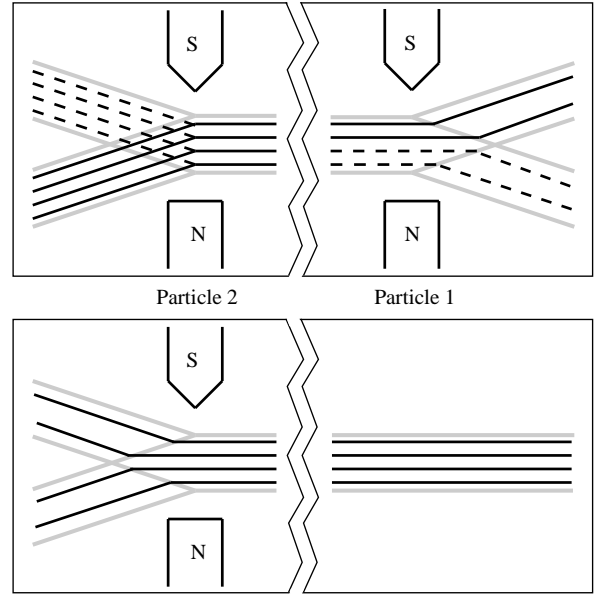
\begin{figure}[t]
\begin{center}
\scalebox{.9}{
\scalebox{1} 
{
\begin{pspicture}(0,-4.46)(8.58,4.46)
\definecolor{color266}{rgb}{0.8,0.8,0.8}
\psline[linewidth=0.04cm](5.5,4.27)(5.5,3.47)
\psline[linewidth=0.04cm](5.5,3.47)(5.9,3.07)
\psline[linewidth=0.04cm](5.9,3.07)(6.3,3.47)
\psline[linewidth=0.04cm](6.3,3.47)(6.3,4.27)
\psline[linewidth=0.04cm](5.5,1.43)(6.3,1.43)
\psline[linewidth=0.04cm](6.3,1.43)(6.3,0.43)
\psline[linewidth=0.04cm](5.5,1.43)(5.5,0.43)
\usefont{T1}{ptm}{m}{n}
\rput(5.91,0.935){$\text{N}$}
\psline[linewidth=0.06cm,linecolor=color266](4.7,2.67)(5.9,2.67)
\psline[linewidth=0.06cm,linecolor=color266](5.9,2.67)(8.3,3.47)
\psline[linewidth=0.06cm,linecolor=color266](5.9,2.67)(8.3,1.87)
\psline[linewidth=0.06cm,linecolor=color266](4.7,1.85)(5.9,1.85)
\psline[linewidth=0.06cm,linecolor=color266](5.9,1.85)(8.3,2.65)
\psline[linewidth=0.06cm,linecolor=color266](5.9,1.85)(8.3,1.05)
\psframe[linewidth=0.04,linecolor=white,dimen=outer,fillstyle=solid](6.16,3.97)(5.66,3.53)
\usefont{T1}{ptm}{m}{n}
\rput(5.88,3.775){$\text{S}$}
\psline[linewidth=0.04cm](4.7,2.57)(6.24,2.57)
\psline[linewidth=0.04cm](4.7,2.37)(6.9,2.37)
\psline[linewidth=0.04cm,linestyle=dashed,dash=0.16cm 0.16cm](4.7,2.17)(6.86,2.17)
\psline[linewidth=0.04cm,linestyle=dashed,dash=0.16cm 0.16cm](4.7,1.97)(6.26,1.97)
\psline[linewidth=0.04cm](6.22,2.57)(8.28,3.27)
\psline[linewidth=0.04cm](6.9,2.37)(8.28,2.87)
\psline[linewidth=0.04cm,linestyle=dashed,dash=0.16cm 0.16cm](6.32,1.95)(8.26,1.31)
\psline[linewidth=0.04cm,linestyle=dashed,dash=0.16cm 0.16cm](6.92,2.13)(8.24,1.65)
\psline[linewidth=0.04cm](3.1,4.25)(3.1,3.45)
\psline[linewidth=0.04cm](3.1,3.45)(2.7,3.05)
\psline[linewidth=0.04cm](2.7,3.05)(2.3,3.45)
\psline[linewidth=0.04cm](2.3,3.45)(2.3,4.25)
\psline[linewidth=0.04cm](3.1,1.43)(2.3,1.43)
\psline[linewidth=0.04cm](2.3,1.43)(2.3,0.43)
\psline[linewidth=0.04cm](3.1,1.43)(3.1,0.43)
\usefont{T1}{ptm}{m}{n}
\rput(2.75,0.935){$\text{N}$}
\psline[linewidth=0.06cm,linecolor=color266](3.9,2.69)(2.7,2.69)
\psline[linewidth=0.06cm,linecolor=color266](2.7,2.69)(0.3,3.49)
\psline[linewidth=0.06cm,linecolor=color266](2.7,2.69)(0.3,1.89)
\psline[linewidth=0.04cm,linestyle=dashed,dash=0.16cm 0.16cm](2.7,2.55)(0.3,3.35)
\psline[linewidth=0.06cm,linecolor=color266](3.9,1.85)(2.7,1.85)
\psline[linewidth=0.06cm,linecolor=color266](2.7,1.85)(0.3,2.65)
\psline[linewidth=0.06cm,linecolor=color266](2.7,1.85)(0.3,1.05)
\psframe[linewidth=0.04,linecolor=white,dimen=outer,fillstyle=solid](2.94,3.95)(2.44,3.51)
\usefont{T1}{ptm}{m}{n}
\rput(2.66,3.755){$\text{S}$}
\psline[linewidth=0.04cm](3.9,2.57)(2.7,2.57)
\psline[linewidth=0.04cm](3.9,2.37)(2.7,2.37)
\psline[linewidth=0.04cm](3.9,2.17)(2.7,2.17)
\psline[linewidth=0.04cm](3.9,1.97)(2.7,1.97)
\psline[linewidth=0.04cm](2.7,2.57)(0.3,1.77)
\psline[linewidth=0.04cm](2.7,2.37)(0.3,1.57)
\psline[linewidth=0.04cm](2.7,2.17)(0.3,1.37)
\psline[linewidth=0.04cm](2.7,1.97)(0.3,1.17)
\psline[linewidth=0.04cm,linestyle=dashed,dash=0.16cm 0.16cm](2.7,2.37)(0.3,3.17)
\psline[linewidth=0.04cm,linestyle=dashed,dash=0.16cm 0.16cm](2.7,2.17)(0.3,2.97)
\psline[linewidth=0.04cm,linestyle=dashed,dash=0.16cm 0.16cm](2.7,1.97)(0.3,2.77)
\usefont{T1}{ptm}{m}{n}
\rput(5.95,0.015){$\text{Particle 1}$}
\usefont{T1}{ptm}{m}{n}
\rput(2.73,0.015){$\text{Particle 2}$}
\psline[linewidth=0.04cm](3.1,-0.43)(3.1,-1.23)
\psline[linewidth=0.04cm](3.1,-1.23)(2.7,-1.63)
\psline[linewidth=0.04cm](2.7,-1.63)(2.3,-1.23)
\psline[linewidth=0.04cm](2.3,-1.23)(2.3,-0.43)
\psline[linewidth=0.04cm](3.1,-3.27)(2.3,-3.27)
\psline[linewidth=0.04cm](2.3,-3.27)(2.3,-4.27)
\psline[linewidth=0.04cm](3.1,-3.27)(3.1,-4.27)
\usefont{T1}{ptm}{m}{n}
\rput(2.75,-3.765){$\text{N}$}
\psline[linewidth=0.06cm,linecolor=color266](3.9,-2.03)(2.7,-2.03)
\psline[linewidth=0.06cm,linecolor=color266](2.7,-2.03)(0.3,-1.23)
\psline[linewidth=0.06cm,linecolor=color266](2.7,-2.03)(0.3,-2.83)
\psline[linewidth=0.06cm,linecolor=color266](3.9,-2.85)(2.7,-2.85)
\psline[linewidth=0.06cm,linecolor=color266](2.7,-2.85)(0.3,-2.05)
\psline[linewidth=0.06cm,linecolor=color266](2.7,-2.85)(0.3,-3.65)
\psframe[linewidth=0.04,linecolor=white,dimen=outer,fillstyle=solid](2.94,-0.73)(2.44,-1.17)
\usefont{T1}{ptm}{m}{n}
\rput(2.66,-0.925){$\text{S}$}
\psline[linewidth=0.04cm](3.9,-2.13)(2.36,-2.13)
\psline[linewidth=0.04cm](3.9,-2.33)(1.7,-2.33)
\psline[linewidth=0.04cm](3.9,-2.53)(1.74,-2.53)
\psline[linewidth=0.04cm](3.9,-2.73)(2.34,-2.73)
\psline[linewidth=0.04cm](2.38,-2.13)(0.32,-1.43)
\psline[linewidth=0.04cm](1.72,-2.33)(0.36,-1.85)
\psline[linewidth=0.04cm](2.34,-2.73)(0.34,-3.39)
\psline[linewidth=0.04cm](1.78,-2.53)(0.36,-3.01)
\psline[linewidth=0.06cm,linecolor=color266](4.72,-2.03)(8.3,-2.03)
\psline[linewidth=0.06cm,linecolor=color266](4.72,-2.85)(8.28,-2.85)
\psline[linewidth=0.04cm](4.72,-2.13)(8.26,-2.13)
\psline[linewidth=0.04cm](4.72,-2.33)(8.26,-2.33)
\psline[linewidth=0.04cm](4.72,-2.53)(8.26,-2.53)
\psline[linewidth=0.04cm](4.72,-2.73)(8.26,-2.73)
\psframe[linewidth=0.02,dimen=outer](8.58,4.37)(0.0,0.31)
\psframe[linewidth=0.02,dimen=outer](8.58,-0.31)(0.0,-4.37)
\psline[linewidth=0.02](4.04,4.37)(4.3,3.77)(4.06,3.37)(4.28,2.77)(4.06,2.35)(4.3,1.73)(4.08,1.35)(4.28,0.77)(4.08,0.33)
\psline[linewidth=0.02](4.26,4.37)(4.52,3.77)(4.28,3.37)(4.5,2.77)(4.28,2.35)(4.52,1.73)(4.3,1.35)(4.5,0.77)(4.3,0.33)
\psline[linewidth=0.02](4.04,-0.31)(4.3,-0.91)(4.06,-1.31)(4.28,-1.91)(4.06,-2.33)(4.3,-2.95)(4.08,-3.33)(4.28,-3.91)(4.08,-4.35)
\psline[linewidth=0.02](4.26,-0.31)(4.52,-0.91)(4.28,-1.31)(4.5,-1.91)(4.28,-2.33)(4.52,-2.95)(4.3,-3.33)(4.5,-3.91)(4.3,-4.35)
\pspolygon[linewidth=0.02,linecolor=white,fillstyle=solid](4.04,4.45)(4.1,4.29)(4.26,4.29)(4.22,4.43)
\pspolygon[linewidth=0.02,linecolor=white,fillstyle=solid](4.3,0.41)(4.24,0.25)(4.04,0.23)(4.16,0.41)
\pspolygon[linewidth=0.02,linecolor=white,fillstyle=solid](4.04,-0.25)(4.1,-0.41)(4.26,-0.41)(4.22,-0.27)
\pspolygon[linewidth=0.02,linecolor=white,fillstyle=solid](4.3,-4.29)(4.24,-4.45)(4.08,-4.45)(4.12,-4.31)
\end{pspicture} 
}
}
\caption{
Illustration of the two scenarios discussed in the text.  In the top
frame, particle 1 (on the right) encounters its $\text{SG}_z$ device
first; the particle is found to be spin-up (solid trajectories) or
spin-down (dashed trajectories) depending on the initial $z$-coordinate of
the particle.  If particle 1 goes up, the collapse suffered
by the CWF of particle 2 (on the left) causes it to go down (solid
trajectories) regardless of its initial $z$-coordinate.  On the other
hand, if particle 1 goes down, the collapse causes particle
2 instead to go up (dashed trajectories) regardless of its initial
$z$-coordinate.  In the lower frame, particle 1 is not subjected to
any measurement.  The result of measuring the $z$-spin of particle 2
is then determined by the initial $z$-coordinate of particle 2.  The
difference between the two scenarios exemplifies the contextuality of
the pilot-wave theory (since the result of measuring the spin of
particle 2 depends not only on the initial state but on whether or not
the spin of particle 1 is measured jointly) and also its non-locality
(since the choice of whether or not to measure particle 1 could be
made at space-like relativistic separation from the measurement of
particle 2).  
\label{fig6}
}
\end{center}
\end{figure}

In order to make the non-local character of the theory absolutely
clear, it is helpful to now consider an alternative scenario in which 
particle 1 is not subjected to any measurement.  For example, perhaps
Alice, who is stationed there next to it, decides (at the last
possible second before particle 1 arrives) to yank the
$\text{SG}_z$ device out of the way.  Then, the two-particle wave
function remains in a state described by Equation \eqref{ent} until
particle 2 arrives at its $\text{SG}_z$ device.  But then particle 2
will behave in just exactly the way we previously described for
particle 1 (when it was the first to encounter an $\text{SG}_z$
device):  its velocity in the overlap region behind the magnets will
be proportional to $\hat{y}$ and the outcome of the spin measurement
will depend on the initial position (in particular, the
$z$-coordinate) of particle 2.  

The non-locality is thus clear:  even for the same initial state -- 
two-particle wave function given by Equation \eqref{ent} and, say, both
particle positions, $\vec{X}_1$ and $\vec{X}_2$, in the upper halves
of 
their respective packets -- the outcome of the measurement on particle 
2 depends on what Alice chooses to do in the
vicinity of particle 1.  If Alice subjects particle 1 to an
$\text{SG}_z$-based measurement, that measurement will have outcome
``spin-up'', the CWF of particle 2 will collapse
to a state proportional to $\chi_{-z}$, and the subsequent measurement
of particle 2's $z$-spin will have outcome ``spin-down''.  On the
other hand, if Alice removes the $\text{SG}_z$ device, the measurement
of particle 2's $z$-spin will instead have the outcome ``spin-up''.
\cite{signaling}
It is clear that the non-locality is a form of contextuality in which
the part of the experimental ``context'' affecting the realized
outcome of the experiment is \emph{remote}.

Although we have focused here on the concrete example in which the
$z$-components of the spins of both particles are (perhaps) measured,
it should now be clear that, and how, the pilot-wave theory accounts
for the empirically observed correlations in the more general sort of
case in which arbitrary components of spin are measured.  One of the
particles will encounter its measuring device \emph{first}; the
outcome of this first measurement will be determined by the random
initial position of this measured particle within its wave packet; the
completion of this first measurement induces a collapse in
the distant particle's CWF; this in turn determines the statistics for
a subsequent measurement on the distant particle.   (Showing that, in
addition, the expected statistical results are reproduced even if one
or both of the measuring devices are replaced with alternative
$\text{SG}'_n$ devices is left as an exercise for the reader.)  

That so much hangs on which measurement happens \emph{first} (or more
mathematically, that the CWF of each particle is the two-particle wave
function evaluated at the actual \emph{current} position of the other
particle, no matter how distant) makes it clear that the non-locality
of the pilot-wave theory will be difficult to reconcile with
relativity.  This is, however, in principle no different from the
non-locality implied by the collapse postulate of ordinary QM.
(Indeed, the collapse of the pilot-wave theory's CWFs can
and should be understood as a mathematically precise \emph{derivation}
of the usual textbook approach to measurement, in a theory where
imprecise notions like ``measurement'' play no fundamental role.)  And
of course, the ``grossly non-local structure'' of the pilot-wave
theory and ordinary QM turns out to be ``characteristic ... of any
such theory which reproduces exactly the quantum mechanical
predictions'' -- as shown in Bell's \emph{other} landmark paper, of
1964.  \cite{bell64}

\section{Conclusions}
\label{sec7}



Contrary to an apparently widespread belief, the pilot-wave theory has
no trouble accounting for phenomena involving spin.  It does so,
actually, in the most straightforward imaginable way:  the wave
function (a scalar-valued field on the configuration space in the case
of spinless particles) is replaced with the appropriate spinor-valued
function, evolving according to the appropriate generalization of the
Schr\"odinger equation, just exactly as in ordinary QM.  The dynamical law for the
motion of the particles, Equation \eqref{guidance}, doesn't change at
all -- one need simply recall, again from ordinary QM, that for
particles with spin the definitions of both $\vec{j}_n$ and $\rho$
involve summing over the spin indices.  Thus understood, the
pilot-wave theory reproduces the usual quantum statistical predictions for
spin measurements (including sequences of measurements performed on a
single particle, and joint measurements performed on pairs of
perhaps-entangled particles) but without any additional postulates or
\emph{ad hoc} exceptions to the usual dynamical laws.  And it does
this, incidentally, in a way that could have been anticipated:  spin
measurements \emph{too} (like position measurements and measurements
whose results are registered by the position of a pointer) ultimately
come down to the position of a particle (downstream from an
appropriate SG device).  

The theory exhibits ``contextuality'' as illustrated in the two
examples discussed:  the result of a measurement  
depends not only on the state of the
particle prior to measurement, but also on the measurement's specific experimental
implementation.  In particular, the outcome of a
measurement of the $z$-component of the spin of a particle can depend
on whether the spin is measured using an $\text{SG}_z$ or an
$\text{SG}'_z$ device, and/or can depend on which (commuting)
observable
is also being jointly measured.  
Considered in the abstract, the idea of a contextual hidden variable
theory perhaps sounds contrived and implausible.  Such, at least, has
apparently been the thinking behind the suggestions that the various
``no hidden variables'' theorems (due to von Neumann, Kochen-Specker,
and so on) rule out the possibility of (or even an important class of)
hidden variables theories.  But the example of the pilot-wave theory shows,
quite simply and conclusively, that they don't.  And
furthermore, like it or lump it,
the theory (backed up by Bell's theorem) 
shows that there is nothing the least bit contrived or
intolerable in the sort of contextuality that is required to eliminate
the ``unprofessionally vague and ambiguous'' aspects of ordinary QM in
favor of something clear and mathematically precise. \cite{vague} 

The idea that there should be something contrived or intolerable about
contextuality undoubtedly arises from the idea that, if a property
really exists, measurement of it should -- by definition -- 
simply reveal its value.  It
would be hard, actually, to disagree with this sentiment.  The key
question, though, is precisely whether any such property exists.  As
has been discussed in illuminating detail in Ref. \cite{nrao},
the real lesson to be taken away from examining the
pilot-wave perspective on spin is that so-called ``contextual
properties'' (like the individual spin components in the pilot-wave
theory) are not properties at all. \cite{history} They simply do not exist and
there is nothing mysterious about this at all, just as there is
nothing mysterious in the fact that the 
eventual flavor of a loaf of bread (which depends not just on
the ingredients but also on how it is later baked!) is not a pre-existing
property of the raw dough:
\begin{quote}
``Note that one can completely understand what's going on in
[a] Stern-Gerlach experiment without invoking any putative property of
the electron such as its actual $z$-component of spin that is supposed
to be revealed in the experiment.  For a general initial wave function
there is no such property.  What is more, the transparency of the [pilot-wave]
analysis of this experiment makes it clear that there is nothing the
least bit remarkable (or for that matter `nonclassical') about the
\emph{nonexistence} of this property.''  \cite{qeop}
\end{quote}
As explained by Bell, the appearance to the contrary -- that is, the
tacit assumption that there \emph{must} be some real ``$z$-component
of spin'' property that the measurements unveil -- seems to arise from
the unfortunate and inappropriate connotations of the word
``measurement'':  
\begin{quote}
``the word comes loaded with meaning from everyday life, meaning which
is entirely inappropriate in the quantum context.  When it is said
that something is `measured' it is difficult not to think of the
result as referring to some pre-existing property of the object in
question.  This is to disregard Bohr's insistence that in quantum
phenomena the apparatus as well as the system is essentially
involved.  If it were not so, how could we understand, for example,
that `measurement' of a component of `angular momentum' -- in an
arbitrarily chosen direction -- yields one of a discrete set of
values?  When one forgets the role of the apparatus, as the word
`measurement' makes all too likely, one despairs of ordinary logic --
hence `quantum logic'.  When one remembers the role of the apparatus,
ordinary logic is just fine.''  \cite{against}
\end{quote}
Unfortunately, even some of the people in the best possible position
to understand this important point -- namely, proponents of the
pilot-wave theory -- have missed it and have thus pointlessly laden the
theory with additional variables
corresponding to such actual spin components.  \cite{holland}  One
goal of the present paper is thus to present, in a clear and
accessible way, the \emph{simplest possible} pilot-wave account of spin, in
order to rectify misconceptions like that held by the anonymous
referee quoted in the introduction.  

Finally, note it is by no means only
champions of hidden variable theories that are sometimes seduced into
thinking of spin components as real properties.
Although the official party line of the orthodox quantum theory, as
expressed for example in standard textbooks, is of course that there
are no such things, one often finds that the authors, perhaps swept
away by all the talk of ``measuring'' ``observables'', are somewhat
conflicted about this.  Townsend, for example, whose beautiful
textbook begins with five full chapters on spin, stresses early on
that, because the operators corresponding to distinct spin components
fail to commute, ``the angular momentum never really `points' in any
definite direction.''  \cite{townsend}  This already implies the
usual, orthodox view that there is simply no such thing as a spin angular
momentum vector, pointing in some particular direction, at all.  

And yet -- even for Townsend, who is \emph{far more careful} about
this issue than most authors \cite{griffiths} -- there is a tendency
to occasionally slip into suggesting that there is, or at least that
(even though there isn't) it is sometimes helpful to pretend that
there is.  Thus for
example, Townsend later explains that  ``placing [a spin-1/2] particle in a
magnetic field in the $z$ direction rotates the spin of the particle
about the $z$ axis as time progresses...''   We also find statements
about how, for example, in the weak decay of the muon, ``the positron
is preferentially emitted in a direction opposite to the spin
direction of the muon'' and even a description of $\langle +x | \psi(t)
\rangle$ and $\langle +y | \psi(t) \rangle$  -- the amplitudes for a
particle in spin state $| \psi(t)\rangle$ to be found, respectively,
spin-up along the $x$- and $y$-directions -- as ``the components of the
intrinsic spin in the $x-y$ plane''.  So perhaps after all particles
\emph{do} have definite spin vectors?  No, Townsend
reminds us that this language and the associated physical picture are not
to be taken too seriously:  
\begin{quote}
``However, we should be careful not to carry over \emph{too} completely the
classical picture of a magnetic moment precessing in a magnetic field
since in the quantum system the angular momentum ... of the particle
cannot actually be pointing in a specific direction because of the
uncertainty relations...''  [emphasis added]
\end{quote}
It is not my intention here to criticize Townsend's text, which I
really do think is wonderful.  Surely it would be made much worse if
all the linguistic shortcuts criticized above were ``fixed''.
(Just imagine the dreariness and impenetrability of a
textbook which explained, for example, that ``the positron is
preferentially emitted in a direction opposite to the direction along
which measurement of the muon's spin, should such a measurement have
been performed prior to its decay, would have given the highest
probability of yielding the outcome `spin-up'.'')  Nevertheless, there
really is a sense -- highlighted especially by Townsend's use of the
word ``too'' in the passage just quoted -- in which the orthodox view
insists on \emph{retaining} the classical picture (of a little
spinning ball of charge with definite spin angular momentum vector)
but simultaneously \emph{apologizing} for this, by demanding that the
picture not be carried over \emph{too} completely, not be taken
\emph{too} seriously.  

The reason for this schizophrenia, I suspect, is that orthodox quantum 
physicsts are, after all, physicists.  They cannot just ``shut up and calculate'' --
not completely.  They need \emph{some} sort of visualizable picture of
what, physically, the mathematical formalism describes, or they simply
cannot keep track of what in the world they are talking about.  \cite{picture}
So they retain the classical picture while simultaneously, out of the other
sides of their mouths, rejecting it.  

Perhaps then, at the end of the day, the most important thing about
the pilot-wave perspective on spin is simply that it provides a
picture of what might actually be going on physically in
phenomena involving spin -- a
picture that does not involve any spinning balls of charge, but which
is nevertheless completely and absolutely \emph{clear} and
\emph{precise} and which \emph{can be taken seriously}, without
apologies or double-speak. 
 
\noindent
\emph{Acknowledgements:} Thanks to George Greenstein for organizing the wonderful series of
meetings at which I had the opportunity to first try out some of this
material on a live audience.  Philip Pearle, John Townsend, Roderich
Tumulka, and two anonymous referees provided helpful  comments on an
earlier draft.


\begin{thebibliography}{99}
\bibitem{field}  Note that this expression for the field is simplified
and idealized:  a more precise expression would
involve field components perpindicular to $\hat{z}$ -- which would
allow the divergence of $\vec{B}$ to vanish, as it of course should 
-- as well as a large
constant field in the $z$-direction.  As it is usually explained, this
large constant field causes rapid precession about the $z$-axis, and
thus makes any transverse forces (that would otherwise arise from the
transverse magnetic field gradients) average to zero.  It is much
simpler, though, to simply look the other way in regard to the
violation of Maxwell's equations, and thus avoid needing to talk about
any of these subtleties, by simply pretending that Equation \eqref{b}
accurately describes the field to which the particles are subjected.
\bibitem{reflectedwave}
If the reflected wave term from Equation \eqref{wf} is retained, the
detailed particle trajectories are a little more complicated.  
During the time period when the incident and reflected
packets overlap, the particle velocity in that ($y<0$) overlap region
is slightly oscillatory, with an average velocity -- in precisely the
sense explained and used in the earlier paper -- that is slightly less than
$\hbar k / m$ by an amount proportional to $(\kappa z)^2$.  Thus,
particles very near the trailing edge of the incident packet will in
fact be overtaken by the trailing edge of the packet before reaching
$y=0$ and will hence be swept away, back out to the left, by the
reflected packet.  In short, particles which happen to begin in a
small (parabolic) slice near the trailing edge of the incident packet
will not behave as described in the main text, but will instead
reflect.  The behavior of the particles which do transmit through to
the $y>0$ region, however, is (in the applicable $\kappa w \rightarrow
0$ limit) unaffected.  So we will continue to suppress discussion at
this level of detail in the main text.
\bibitem{angle}  Actually there is one subtle point that deserves to
  be addressed.  In analyzing the measurement of $z$-spin, we have
  assumed that the incident wave packet has an essentially rectangular
  cross-sectional profile, with (as mentioned explicitly) some width
  $w$ along the $z$-direction and (as only really assumed tacitly)
  some width $w'$ along the $x$-direction.  This makes the problem
  ``effectively 2D'' in the sense that (for the width-$w'$ range of
  $x$-values where the wave function has support and where the
  particle might therefore be) nothing depends on $x$.  If, however,
  the incident wave function is the same but the 
  SG apparatus is rotated through some arbitrary angle (not an integer 
  multiple of $90^\circ$), this nice symmetry is spoiled.  Everything
  still works as claimed; it's just that one has to think a little bit
  (about, e.g., slicing the 3D wave packet up into a stack of planes
  perpindicular to $\hat{n}$) to convince oneself of this.
  Alternatively, one can recognize that the statistics of the
  measurement outcomes will not depend on the exact cross-sectional
  profile of the incident packet and hence imagine that the
  cross-sectional profile of the incident packet is rotated in tandem
  with the SG device so that the analysis remains transparent.
\bibitem{future}  Of course, the mere fact that the two sub-beams are
  spatially-separated is not sufficient to prove that the empty packet
  is forevermore dynamically irrelevant to the motion of the
  particle.  The ``empty'' spin-down sub-beam
  could be made to become, again, dynamically relevant -- for example, 
  by arranging, by use (say) of
  appropriate magnetic fields, to deflect the spin-down beam back up
  toward the spin-up beam, at which point interference effects
  (including perhaps the re-constitution of a guiding wave
  proportional to $\binom{1}{0}$!) could occur.  But if, as suggested in Figure \ref{fig4},
  the spin-down sub-beam is blocked by a brick wall, then the usual
  process of decoherence will render it impossible, at least for all
  practical purposes, to arrange the appropriate overlapping of the
  two components of the wave function, and treating the wave as having
  ``effectively collapsed'' is entirely warranted.
\bibitem{delta}  We assume here that $\Delta > w$ so that the two
  sub-beams downstream of the SG device do not overlap spatially.
\bibitem{signaling}  It is perhaps worth noting explicitly that the
  non-locality cannot be used for signaling:  if a large ensemble of
  pairs are prepared in the state \eqref{ent}, and Alice chooses at
  the last minute whether to subject all of the particles on her side
  to an $\text{SG}_z$ measurement, or some other spin measurement, or
  no measurement, Bob will see the same $50/50$ statistics on his side
  regardless of her choice.
\bibitem{picture}  In
this respect it is somewhat telling that the worst offenses against
the official orthodox view tend to occur in figures and
illustrations.  Almost all books for example
include diagrams depicting the precession, induced by the presence of
a magnetic field, of a particle's spin vector about some axis.  Townsend's text
avoids that particular misleading suggestion of a classical picture.
But his front cover art -- depicting a Stern-Gerlach spin measurement
much like our Figure \ref{fig1} -- includes little circles with arrows
(pointing respectively up and down) in the downstream sub-beams.  The
circles are even yellow, inviting us to recall Bell's warning, quoted
earlier in note \cite{history}.
\end{thebibliography}
\end{document}